\documentclass[10pt]{article}
\usepackage{graphicx,floatflt,amssymb,epsfig}
\newcommand{\vtwographs}[2]{%
\unitlength=1in
\begin{picture}(4,6)
\put(0,0){\epsfig{file=#2.ps, width=4in}}
\put(0,3){\epsfig{file=#1.ps, width=4in}}
\put(-0.2,5){(a)}
\put(-0.2,2){(b)}
\end{picture}}

\begin{document}
\begin{flushright}
\texttt{hep-ph/0408295}\\
SINP/TNP/04-14\\
TIFR/TH/04-20\\
\end{flushright}

\vskip 50pt

\begin{center}
{\Large  \bf  Searching for signals of minimal length in extra dimensional
models using dilepton production at hadron colliders} \\
\vspace*{1cm} \renewcommand{\thefootnote}{\fnsymbol{footnote}} {\large
{\sf Gautam Bhattacharyya ${}^1$},  {\sf Prakash Mathews ${}^1$}, {\sf
Kumar Rao ${}^2$}, and {\sf K. Sridhar ${}^2$} } \\
\vspace{10pt}

{\small  1. Saha  Institute  of Nuclear  Physics,  1/AF Bidhan  Nagar,
        Kolkata 700064, India \\ 2. Department of Theoretical Physics,
        Tata  Institute of Fundamental  Research, \\Homi  Bhabha Road,
        Mumbai 400005, India }

\normalsize
\end{center}

\begin{abstract}

\noindent Theories of quantum gravity suggest the existence of a
minimal length scale. It is interesting to speculate what the
consequences of the existence of such a length scale would be for
models of large extra dimensions, in particular, the ADD model. When
ADD model is conflated with the minimal length scale scenario,
processes involving virtual exhange of gravitons cease to be
ulraviolent divergent. We study the production of dileptons at hadron
colliders as an example of a process mediated by virtual gravitons. We
find that the bounds we derive on the effective string scale are
significantly different (in fact, less stringent) from those derived
in the conventional ADD model, both at the upgraded Tevatron and the
Large Hadron Collider.

\vskip 5pt \noindent
\texttt{PACS Nos:~ 11.25.Wx, 13.85.Qk} \\
\texttt{Key Words:~~Dilepton production at hadron colliders, Extra 
dimensions, Minimal length.}
\end{abstract}

\renewcommand{\thesection}{\Roman{section}}
\setcounter{footnote}{0}
\renewcommand{\thefootnote}{\arabic{footnote}}

\section{Introduction}
A reinterpretation of the gauge hierarchy problem by Arkani-Hamed,
Dimopolous and Dvali (ADD) \cite{lxd} involves an alteration in the
behaviour of gravity at small distances owing to the existence of
large extra compactified space dimensions.  In the ADD scenario, the
gauge interactions are confined to a 3-brane while gravity propagates
in all the dimensions. The effective Planck scale $M_S$ in this model
gets related to the usual Planck scale by $M_{\rm{Pl}}^{2}= R^{d}
M_{S}^{d+2}$, where $R$ is the radius of compactification and $d$ is
the number of extra dimensions.  It is possible to choose $R$ for a
given $d$ such that the fundamental string scale comes down to $M_S
\sim$ 1 TeV.  Such a low value of $M_S$ is also exciting from the
point of view of studying quantum gravity at present and future
colliders. Various signals of graviton production and virtual graviton
exchange have been proposed in several papers in the last few years
\cite{hlz, grw}. These are also reviewed in Refs.~\cite{perez, self}.

Another important general feature of quantum gravity theories, apart
from the requirement of extra dimensions, is the existence of a
minimal length scale (MLS). In string theory, such a minimal length is
suggested since strings cannot probe distances smaller than the string
scale. If the energy of a string reaches the Planck scale, string
excitations can occur causing its extension \cite{witten}.  Even
though we are inspired by this assertion of quantum gravity theories,
nevertheless we implement the notion of a minimal length scale
phenomenologically. Note that in some fundamental theory, like string
theory, such a notion may have a different meaning and
interpretation. In the context of large extra dimensional models
exhibiting a low fundamental scale of $M_S$ of order 1 TeV, the
existence of a minimal length becomes phenomenologically important if
we take it to be around an inverse TeV, viz. $l_{p} \sim 1/M_S$. In
the ADD model, for an amplitude involving virtual gravitons, one has
to sum over an infinite tower of graviton Kaluza-Klein (KK)
states. The result is divergent and to cure the divergence an {\it
ad-hoc} cutoff of the order of $M_{S}$ is used, see Ref. \cite{hlz}
for details. A very attractive feature of the MLS scenario is that one
can sum over the entire KK graviton tower because the contribution of
higher energy KK states is smoothly cut off, as we shall see later,
rendering the amplitude finite. This constitutes the main motivation
for incorporating the effects of a minimal length scale in the ADD
model. This also leads to a significant deviation of the bounds on the
conventional ADD model parameters, e.g. $M_S$, derived from collider
searches, as we shall see later.

\section{The MLS scenario and $\mathbf{q\bar q \to l^+l^-}$}
We define the MLS scenario to mean the ADD model with the idea of a
minimal length $l_p$ incorporated in it with $l_p$ to be of the order
of TeV${}^{-1}$. The uncertainty in position measurement now cannot be
smaller than $l_p$, which means that the standard commutation
relation between position and momentum needs to be modified
\cite{kempf, hassan}. In the minimal length scheme, since distances
less than $l_{p}$ do not exist, the Compton wavelength ($\lambda =
2\pi/k$) of a particle cannot be arbitrarily small.  However, we
suppose that even though the wave vector $k$ is bounded from above,
the momentum $p$ can be made as large as possible.  Similarly, to
maintain the same dispersion relation the frequency $\omega$ is
restricted from above while the energy $E$ can go up arbitrarily. This
immediately requires that the standard relations $p=\hbar k$ and
$E=\hbar \omega$ have to be modified. This can be realised by
introducing the following ansatz for the modified relations, known as
the Unruh relations \cite{unruh}
\begin{eqnarray}
\label{dispersion}
l_{p}k(p)&=& \tanh^{1/\gamma}
\left[\left(\frac{p}{M_{S}}\right)^{\gamma}\right],  \nonumber \\ 
l_{p} \omega(E)&=&\tanh^{1/\gamma}
\left[\left(\frac{E}{M_{S}}\right)^{\gamma}\right],
\end{eqnarray}
where $\gamma$ is a positive constant. Hereafter, we take $\gamma=1$
for simplicity and set $l_{p} M_{S} = \hbar = 1$.  The above simple
relations capture the essence of a minimal length scale: as $p$ (or
$E$) becomes very large, $k$ (or $\omega$) approaches the upper bound
$1/l_{p}$ and hence one cannot probe arbitrarily small length and time
scales.  The relations (\ref{dispersion}) lead to a generalized
position-momentum and energy-time uncertainty principle which can be
written in a Lorentz covariant form as
\begin{equation}
\label{ur}
\left[x^{\nu},  p_{\mu}\right]=i  \,\,\frac{\partial p_{\mu}}{\partial
k_{\nu}}. 
\end{equation}
As explained in \cite{kempf,hossen1}, the effect of the modified
relations (\ref{dispersion}) can conveniently be accounted for by a
simple redefinition of the momentum measure, which in one dimension is
\begin{equation}
dp \longrightarrow dp \,\,\frac{\partial k}{\partial p}. 
\end{equation}
The Lorentz invariant 4-momentum measure scales as
\begin{equation}
\label{scale}
d^{4}p    \to     d^{4}p    \,\,\,\textrm{det}    \left(\frac{\partial
k_{\mu}}{\partial  p_\nu}\right)=d^{4}p  \prod  _{\nu}  \frac{\partial
k_{\nu}}{\partial p_\nu}, 
\end{equation}
where the Jacobian matrix $\frac{\partial k_{\nu}}{\partial p_{\nu}}$
can be kept diagonal.

Some applications of the minimal length scenario have been discussed
in Refs.~\cite{harbach, hossen2}.  In this paper, we study dilepton
production at hadron colliders as a probe of extra dimensional models
with a minimal length scale. To this end, we compute the dilepton
production cross section at hadron colliders (i) in the SM (the
standard Drell-Yan process), (ii) in the conventional ADD model (i.e.,
with no minimal length), and (iii) in the MLS scenario.  In the ADD
model and in the MLS scenario, apart from the SM diagrams, the
dilepton production cross section receives contributions from diagrams
involving virtual gravitons coupled to quarks and gluons. As mentioned
previously, for diagrams involving virtual gravitons the amplitude is
a sum over an infinite tower of KK states having masses $m_{n}=n/R$,
where $n$ is an integer.  Each state couples to fermions with a
strength $1/M_{\rm Pl}$ and a summation over the KK states enhances
the effective strength to $1/M_{S}$. The derivation of the Feynman
rules for the ADD model can be found in \cite{hlz} and the expression
for the Drell-Yan differential cross section is displayed in
\cite{cheung}. In the expression for the squared matrix-element the
interference term between the SM diagram and the diagram with graviton
exchange goes like $F/M_{S}^4$, while the pure graviton term goes like
$(F/M_{S}^4)^2$. In Appendix B of Ref.~\cite{hlz}, an expression for
$F/M_{S}^4$ has been derived after summing over the graviton KK tower:
\begin{eqnarray} 
\label{fadd}
\frac{F}{M_S^4} = \frac{-2}{M_S^{d+2}} (\sqrt{\hat{s}})^{d-2}
I(M_S/\sqrt{\hat{s}}), 
~~~{\rm where}~~~  I(\Lambda) = P \int_0^{\Lambda} dy
\frac{y^{d-1}}{1-y^2},
\end{eqnarray}   
where $\sqrt{\hat{s}}$ is the partonic center-of-mass energy.  The
integral $I(\Lambda)$ is ultraviolet divergent, showing that the
summation over the graviton KK states is infinite, as mentioned
before. But in the MLS scenario, because of the Unruh relations, the
integral is modulated by a factor $\partial \omega/\partial E$ which
smoothly cuts off the contribution of the higher energy KK
states. Hence we can perform the integration over all the KK
states\footnote{While computing the {\em pure} graviton term in the
cross section, we also include the resonant contribution by actually
taking $\sqrt{I'^2+\pi^2/4}$ in place of $I'$ in Eq.~(\ref{fmls}).}:
\begin{eqnarray} 
\label{fmls}
\frac{F}{M_S^4} \to \frac{F'}{M_S^4} = \frac{-2}{M_S^4}
\left(\frac{\sqrt{\hat{s}}}{M_S}\right)^{d-2}  I', 
~~~{\rm  where}~~~
I' = P \int_0^\infty dy \frac{y^{d-1}}{1-y^2}~{\rm
sech}^2\left(y\frac{\sqrt{\hat{s}}}{M_S}\right).
\end{eqnarray}
The sech-square function for large $y$ goes like exp($-2y$) which
damps the power law growth of $y$, thus avoiding the need for an {\em
ad-hoc} cutoff as in $I$. The divergence is thus dynamically remedied
by the requirement of minimal length.

Another important modification due to the MLS is the change in the
cross section due to the rescaling of the momentum measure given in
Eq.~(\ref{scale}). It is straightforward to check, as derived in
\cite{hossen1}, that the phase space integration in the total cross
section picks up the following modification factor:
\begin{equation} 
\label{mod}
d\sigma ({\rm modified}) = d\sigma~\prod_n \frac{E_n}{\omega_n} ~
\prod_\nu  \left. \frac{\partial k_\nu}{\partial  p_\nu}\right|_{p_i =
p_f},
\end{equation}
where $n$ runs over the four initial and final states in a $2 \to 2$
process, and $p_i$ and $p_f$ are the total initial and final four
momenta in a $2 \to 2$ process.  We work out this modification factor
for the process we are studying viz. $ab \rightarrow l^+ l^-$, where
$a,\ b$ are the initial state partons.  Using Eqs. (\ref{dispersion})
and (\ref{scale}), we can easily show that
\begin{eqnarray} 
\label{ebyomega}
\prod_n \frac{E_n}{\omega_n} & = & \frac{s p_T^2}{4M_S^4} \prod_{i=1,2}
x_i ~\cosh(y_i) ~\textrm{coth}\left(\frac{x_i\sqrt{s}}{2M_S}\right)
~\textrm{coth}\left\{\frac{p_{T}\cosh(y_i)}{M_S}\right\}, \\ 
\prod_\nu
\frac{\partial k_\nu}{\partial p_\nu} & = & {\rm   sech}^2
\left\{\frac{(x_1+x_2)\sqrt{s}}{2M_S} \right\} {\rm sech}^2
\left\{\frac{(x_1-x_2)\sqrt{s}}{2M_S} \right\},
\end{eqnarray} 
where $x_1$ and $x_2$ are the momentum fractions of the hadrons
carried by the initial state partons, while $y_1$ and $y_2$ are the
pseudorapidities and $p_T$ is the common transverse momentum of the
final state leptons.  It can be easily checked that in the decoupling
limit $M_S \gg \sqrt{s}$, the phase space correction factor goes to
unity.

\section{Discussion of results}
In \cite{cheung}, the effect of virtual graviton exchange has been
studied for dilepton production in the conventional ADD model. A lower
bound of $M_S$ in the 1 to 3.5 TeV range (Tevatron Run I and II) and
6.5 to 12.8 TeV range (LHC) has been obtained at 95\% C.L. by varying
$d$ in the range 2 to 7. To incorporate the effects of the minimal
length, we modify the expression for the differential cross section
for the Drell-Yan process given in \cite{cheung} according to
Eqs.~(\ref{fmls}) and (\ref{mod}).

\begin{figure}%
\vtwographs{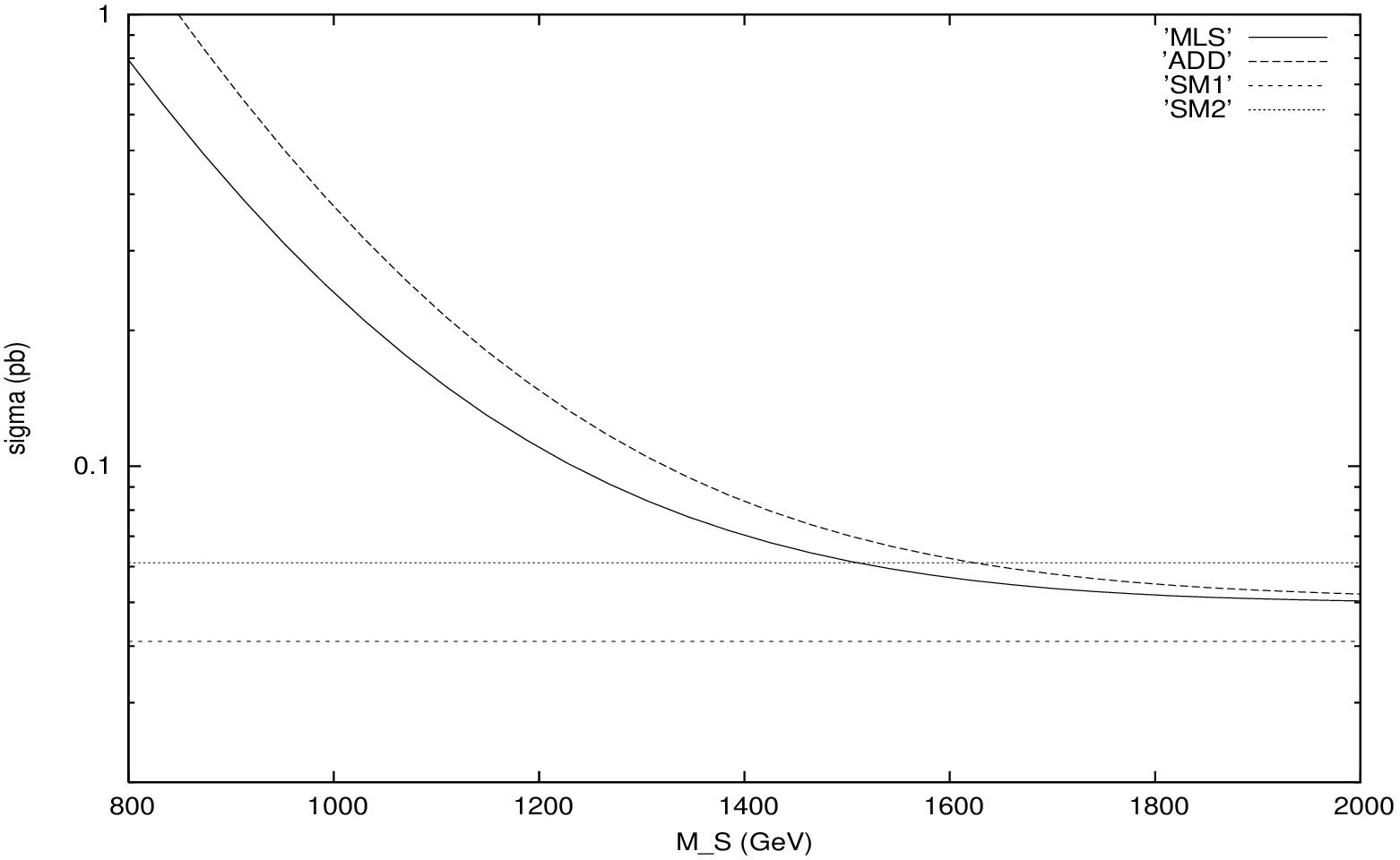}{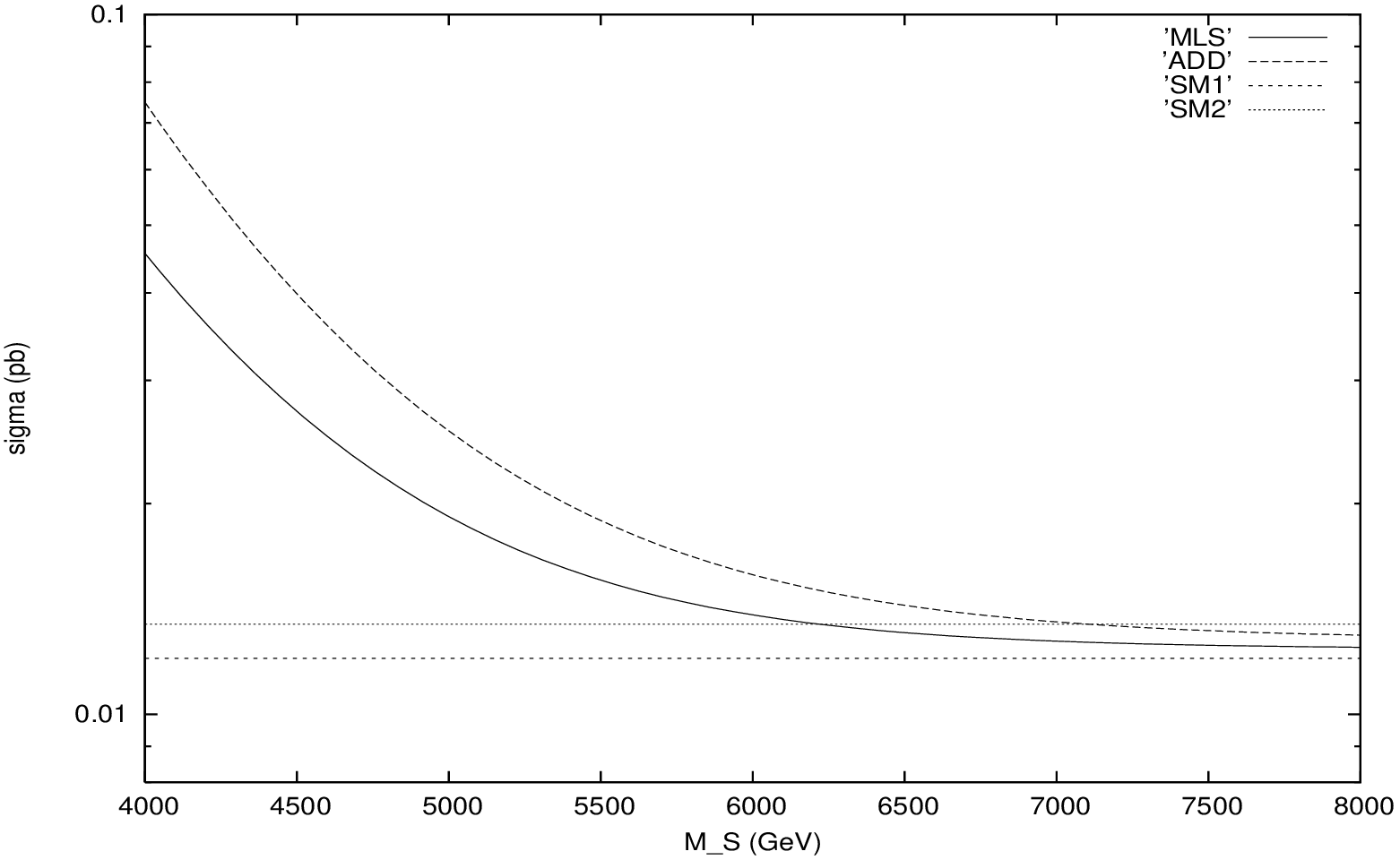}
\caption{The dilepton cross section integrated over invariant mass $M$ of the
dilepton pair as a function of the fundamental scale $M_S$ at the (a) Tevatron
Run II with $M > 250$ GeV and, (b) LHC with $M >600$ GeV. The two curves are
for the conventional ADD model and the MLS scenario, assuming the number of
extra dimensions $d$ to be 3.  Also shown are the 95\% C.L. upper and lower
bounds (`SM2' and `SM1', respectively) on the SM cross section (assuming only
statistical errors).}
\label{xsec}
\end{figure}

Fig.~1 contains the results of our numerical computation. Fig (1a) shows the
dilepton production cross section, integrated over dilepton invariant masses
greater than 250 GeV and over the pseudorapidity range $|y_1|, |y_2| \le 1.1$
and $1.5 < y_1, y_2 < 2.5$, as a function of the scale $M_{S}$ for $p\bar{p}$
collisions at the upgraded Tevatron (Run II) operating at an energy of
$\sqrt{s}=2$ TeV and an integrated luminosity of 2 $\textrm{fb}^{-1}$. We have
used the MRS2001 LO parton densities\footnote{In principle, the MLS hypothesis
should also modify the parton densities. But the effects are numerically
insignificant: the parton densities are calculated by employing the QCD input
data obtained at a few GeV scale, while the inverse minimal length is a
TeV. The $Q^2$ evolution of the parton densities is also not numerically
sensitive to this issue. A similar conclusion has been drawn in \cite{hossen3}
in the context of black hole production in hadron colliders. We thank
S.~Hossenfelder for bringing this point and Ref.~\cite{hossen3} to our
attention.}. The two curves shown in Fig.~(1a) are for the conventional ADD
model\footnote{Our treatment of the ADD model is somewhat different from that
presented in Ref.~\cite{cheung}. While the authors of Ref.~\cite{cheung} have
approximated the expression for $F$ in Eq.~(\ref{fadd}) by the leading terms
yielding $F=\textrm{ln}(M_{S}^2/\hat{s})$ for $d=2$ and $F=2/(d-2)$ for $d>2$,
we have used the full expressions for the integral appearing in
Eq.~(\ref{fadd}).} and MLS scenario respectively. We have taken $d=3$ for the
curves shown in the figure\footnote{Since our primary aim is to compare the
bounds obtained in the conventional ADD model and the MLS scenario, we have
restricted ourselves to a fixed value of $d = 3$. We note that the $d$
dependence appears in the graviton summation and in the MLS scenario the
sech-square modulation of the integral in Eq.~(\ref{fmls}) renders that
dependence rather weak.}.  We have also shown the 2-$\sigma$ upper and lower
limits of the SM, assuming only statistical errors. We observe that the cross
section for the MLS is smaller than that for ADD. This is mainly because of
the exponential suppression of the phase space factor.  For the ADD case, a
bound of about 1.63 TeV results at the 95\% C.L., but this bound is diluted in
the MLS scenario where we get a 95\% bound of about 1.45 TeV. Fig.~(1b)
displays the case of LHC ($pp$ collisions at $\sqrt{s}=14$ TeV), where we
assume a luminosity of 100 $\textrm{fb}^{-1}$ and integrate over dilepton
invariant masses greater than 600 GeV and over the pseudorapidity range
$|y_1|, |y_2| \le 2.5$. Again, we have taken $d=3$.  We obtain the lower bound
on $M_S$ to be about 7 TeV for the conventional ADD model which goes down to
about 6 TeV for the MLS scenario.

\section{Conclusions}
A minimal length scale is a generic prediction of quantum gravity
theories. In brane world models, like the ADD model, the implications
of such a minimal length can be probed in collider
experiments in the TeV range. We study dilepton production at hadron
colliders mediated by virtual graviton exchange in an ADD model having
a minimal length of the order of the inverse of the effective string
scale.  The technical benefit of introducing the minimal length into
the ADD model is that the sum over the graviton tower is regulated
which does not any more require the introduction of an {\em ad-hoc}
cutoff. We find that the bounds are substantially lowered in such a
scenario as compared to the conventional ADD model without the minimal
length scale. Our choice of studying the dilepton production cross
section in hadron colliders is an illustrative one; similar analyses
can be carried out for other processes. In all cases, the suppression
factors associated with minimal length would relax the existing
constraints.

\section*{Acknowledgements}
G.B. and K.S. have benefitted from conversations with Emilian Dudas
and they wish to thank him for the discussions.  G.B., P.M. and
K.S. acknowledge support by the Indo-French Centre for the Promotion
of Advanced Research, New Delhi, India (IFCPAR Project No:
2904-2). G.B.'s research has also been supported, in part, by the DST,
India, Project No: SP/S2/K-10/2001.

\end{document}